\documentclass[superscriptaddress,twocolumn,longbibliography,amsmath,amssymb,floatfix,aps]{revtex4-1}
\usepackage[]{graphicx,color}
\usepackage{amsmath}
\usepackage{algorithm}
\usepackage{algcompatible}
\usepackage{color}
\usepackage{mathtools}
\usepackage[utf8]{inputenc}
\usepackage{CJKutf8}
\usepackage{xspace}

\definecolor{darkgreen}{rgb}{0,0.5,0}

\begin{document}
\title{
Spin-wave dispersion and exchange stiffness in 
Nd$_2$Fe$_{14}$B
and
$R$Fe$_{11}$Ti ($R$=Y, Nd, Sm)
from
first-principles calculations
}
\author{Taro \surname{Fukazawa}}
\email[E-mail: ]{taro.fukazawa@aist.go.jp}
\affiliation{CD-FMat, National Institute of Advanced Industrial Science
and Technology, Tsukuba, Ibaraki 305-8568, Japan}
\affiliation{ESICMM, National Institute for Materials Science,
Tsukuba, Ibaraki 305-0047, Japan}

\author{Hisazumi \surname{Akai}}
\affiliation{The Institute for Solid State Physics, The University of Tokyo,
Kashiwa, Chiba 277-8581, Japan}
\affiliation{ESICMM, National Institute for Materials Science,
Tsukuba, Ibaraki 305-0047, Japan}

\author{Yosuke \surname{Harashima}}
\affiliation{CD-FMat, National Institute of Advanced Industrial Science
and Technology, Tsukuba, Ibaraki 305-8568, Japan}
\affiliation{ESICMM, National Institute for Materials Science,
Tsukuba, Ibaraki 305-0047, Japan}
\affiliation{Institute of Materials and Systems for Sustainability,
Nagoya University, Nagoya, Aichi 464-8601, Japan}

\author{Takashi \surname{Miyake}}
\affiliation{CD-FMat, National Institute of Advanced Industrial Science
and Technology, Tsukuba, Ibaraki 305-8568, Japan}
\affiliation{ESICMM, National Institute for Materials Science,
Tsukuba, Ibaraki 305-0047, Japan}

\date{\today}
\begin{abstract}
We theoretically investigate spin-wave dispersion in rare-earth magnet compounds 
by using first-principles calculations
and a method we call the reciprocal-space algorithm (RSA). 
The value of the calculated exchange stiffness for Nd$_2$Fe$_{14}$B 
is within the range of reported experimental values. 
We find that the exchange stiffness is considerably anisotropic when only short-range exchange couplings are considered, 
whereas inclusion of long-range couplings weakens the anisotropy. 
In contrast, $R$Fe$_{11}$Ti ($R$=Y, Nd, Sm) shows large anisotropy in the exchange stiffness. 
\end{abstract}
\preprint{Ver. 3.0.4}
\maketitle
\section{Introduction}
Finite-temperature properties of magnetic materials are one of 
the central interests in the development of permanent magnets 
not only from a fundamental point of view but also from a technological point of view,
because magnets are used at and above room temperature in industrial applications. 
Modern high-performance permanent magnets are rare-earth magnets, and their main phases are rare-earth transition-metal compounds. 
The magnet compound Nd$_2$Fe$_{14}$B is
the main phase of the neodymium magnet,
which is often said to be the strongest magnet.
$R$Fe$_{12}$-type compounds with 
a ThMn$_{12}$-type structure\cite{Hirayama17} 
have also attracted attention
as potential main phases of magnets for surpassing
the neodymium magnet.

Spin-wave frequency is a fundamental quantity 
for studying magnetic properties at finite temperatures. 
It is a collective and elementary excitation in magnetic materials. 
The curvature of its lowest branch at the $\Gamma$ point
is called the spin-wave stiffness 
and is proportional to the exchange stiffness. 
Experimentally, the spin-wave stiffness can be measured
directly by neutron scattering
or by deduction from the magnetization curve versus temperature
on basis of Bloch's $T^{3/2}$ law.

For example, the exchange stiffness of Nd$_{2}$Fe$_{14}$B
as determined by neutron scattering experiments was 
reported by Mayer et al.\cite{Mayer91,Mayer92} and by Ono et al.\cite{Ono14}
In addition, inelastic neutron scattering was recently reported in a study by 
Naser et al.\cite{Naser20} 
Although Mayer et al. observed no significant difference
in the spin-wave dispersion along the $a$ direction and the $c$ direction,
Naser et al. argued that 
the exchange stiffness has significant anisotropy below room temperature, 
where the stiffness in the $c$ direction ($A_z$) is smaller
than in the $a$ direction ($A_x$). 

From a theoretical point of view, 
the exchange stiffness is a basic input parameter in micromagnetic simulations. 
Although experimental values are used in most simulations, 
recent efforts have been made to determine the exchange stiffness theoretically based on first-principles calculations.
Toga et al. derived a classical Heisenberg model of Nd$_{2}$Fe$_{14}$B 
from the results of density functional calculations.\cite{Toga18} 
The model thus obtained was analyzed by Monte Carlo simulation to determine
the exchange stiffness at finite temperatures.
The exchange stiffness was found to have significant anisotropy 
with $A_z < A_x$. Gong et al. also performed a similar simulation 
and obtained a qualitatively similar result.\cite{Gong19,Gong20}
However, Toga et al. noted
that the anisotropy was greatly weakened when they considered more
interactions between spins, which is
examined in the present paper.

Previous reports on the exchange stiffness in Nd$_{2}$Fe$_{14}$B have
significant variation,
ranging from 6.6 pJ/m \cite{Ono14} to 18 pJ/m \cite{Naser20}.
A direct approach to spin-wave dispersion
from theory can offer key information for resolving the problem
by estimating 
the exchange stiffness from the spin-wave stiffness. 
There are several schemes based on 
first-principles calculations for 
spin-waves, such as those with frozen magnon approaches 
\cite{Halilov98,VanSchilfgaarde99,Shallcross05,Yu08},
those with perturbational approaches based on multiple scattering
theory\cite{Oguchi83,Liechtenstein87,Mryasov96,Pajda01,Sipr19}, and 
those based on direct calculations of
magnetic susceptibility\cite{Okumura19}.
However, these methods are computationally heavy
and do not seem to be practical for 
systems containing more than a few atoms in the unit cell.
Pajda et al.\cite{Pajda01} pointed out that
even for bcc Fe, fcc Co, and fcc Ni, 
calculation of the spin-wave stiffness is 
time-consuming and convergence with respect to
the spatial cutoff for bond lengths is very slow.

To resolve these problems,
we recently developed the reciprocal-space algorithm
(RSA), an accelerated computational
scheme for spin-wave dispersion \cite{Fukazawa19} based on
Liechtenstein's method.\cite{Liechtenstein87}
The scheme completes calculation of the dispersion
without performing inverse Fourier transform to obtain
the spatial representation of magnetic interactions.
This optimization makes the scheme much faster,
and allows it to deal with larger systems than previous methods can handle.

In the present work, we apply the scheme to 
Nd$_2$Fe$_{14}$B and 
$R$Fe$_{11}$Ti ($R$ = Y, Nd, Sm). 
As mentioned above, the anisotropy
in the exchange stiffness has been a topic of focus
in many publications.\cite{Toga18,Gong20,Naser20}
Our results suggest that the anisotropy in Nd$_2$Fe$_{14}$B is
much smaller than previous results, and that it is rather isotropic. 
In contrast,
RFe$_{11}$Ti exhibits significant anisotropy in its exchange stiffness
in our results. 
We discuss this difference in anisotropy 
between Nd$_2$Fe$_{14}$B and a ThMn$_{12}$-type compound
in relation to differences in the unit cell. 
We also show the angular integrated spin-wave
dispersion ($E$--$|q|$ dispersion)
for comparison with
diffraction experiments with polycrystalline and
powder samples. 

\section{Methodology}
\label{S:Methodology}
We perform first-principles calculations
based on
density functional theory
within the local density approximation 
\cite{Hohenberg64,kohn65}.
We use the Korringa-Kohn-Rostoker (KKR) Green function method. 
The spin-orbit coupling (SOC) is not included explicitly, 
but SOC for f-electrons at the rare-earth sites is implicitly considered
as a trivalent open core obeying Hund's rules.
The self-interaction correction\cite{Perdew81} is also applied
to those f-electrons.

We calculate
magnetic couplings
according to Liechtenstein's formula,\cite{Liechtenstein87}
in which the coupling $J_{(i,a)(j,b)}$ between the $i$th site
in the $a$th cell
and the $j$th site in the $b$th cell
is written as 
\begin{equation}
  J_{(i,a)(j,b)}
  =
  \frac{1}{4\pi}
  \Im
  \int^{\epsilon_\mathrm{F}}_{-\infty}
  d\epsilon\,
  \mathrm{Tr}\,
  \Delta_i
  T^{(i,a)(j,b)}_{\uparrow}
  \Delta_j
  T^{(j,b)(i,a)}_{\downarrow},
  \label{Eq:Liechtenstein_Formula}
\end{equation}
where $\epsilon_\mathrm{F}$ denotes the Fermi energy,
$T^{(i,a)(j,b)}_\sigma$ is the scattering-path operator of
the $\sigma$-spin electron from the $(i,a)$ site
to the $(j,b)$ site, and
$\Delta_i$ describes the spin-rotational perturbation
and is defined as
$\Delta_i = t^{-1}_{i,\uparrow} - t^{-1}_{i,\downarrow}$
where $t_{i,\sigma}$ denotes
the t-matrix of the $\sigma$-spin
potential of the $i$th site.
Although
the variables $\Delta_i$ and $T^{(i,a)(j,b)}_{\sigma}$ are functions
of the angular momenta and energy, for simplicity we omitted 
explicitly showing these dependencies.
The trace is taken with respect to the angular momenta,
and the integral is with respect to the energy.

As we describe below,
the Fourier transform
of $J_{(i,a)(j,b)}$ is needed for calculating the spin-wave dispersion:
\begin{equation}
  J_{i,j}(\vec{q})
  =
  \frac{1}{N}
  \sum_a
  J_{(i,0)(j,a)}
  e^{-i\vec{q}\cdot (\vec{R}_a-\vec{R}_0)},
\end{equation}
where $\vec{R}_a$ denotes the position of the $a$th cell.
We directly calculate $J_{i,j}(\vec{q})$
in reciprocal space.
This benefits the precision because
the KKR scheme involves calculation of
the scattering-path operator in reciprocal space,
and transformation of the operator into real space
introduces additional errors.
However, the perturbation $\Delta_i$
is a $\delta$-like local quantity, and can 
be transformed to reciprocal space
without any loss of precision:
$\Delta_i(\vec{q})=\Delta_i$.
The transformed expression of Eq. \eqref{Eq:Liechtenstein_Formula}
is as follows:
\begin{equation}
  J_{i,j}(\vec{q})
  =
  \frac{1}{4\pi}
  \Im
  \sum_{\vec{k}}
  \int^{\epsilon_\mathrm{F}}_{-\infty}
  d\epsilon\,
  \mathrm{Tr}\,
  \Delta_i
  T^{i,j}_{\uparrow}(\vec{k}+\vec{q})
  \Delta_j
  T^{j,i}_{\downarrow}(\vec{k}).
  \label{Eq:Liechtenstein_Formula_in_q}
\end{equation}

We can construct
the classical Heisenberg model by using the values of $J_{(i,a),(j,b)}$
as follows:
\begin{equation}
  H = -\sum_{l=(a,i)} \sum_{\substack{m=(b,j) \\ m\neq l}} J_{lm}\,
  \vec{e}_{l} \cdot \vec{e}_{m},
\end{equation}
where $\vec{e}$ denotes
the unit vector that indicates the direction of the spin moment of the site.
We fix the spin moment $S_i$ to
the value at the ground state, and thus the spin vector is $\vec{S}_{(a,i)} = S_i \vec{e}_{(a,i)}$.
We renormalize
the magnetic coupling into the form 
$\tilde{J}_{(a,i)(b,j)} = J_{(a,i)(b,j)}/(S_i S_j)$.
As a result, the Hamiltonian is cast into the form 
\begin{equation}
  H =
  -\sum_{l=(a,i)} \sum_{m=(b,j)}
  \tilde{J}_{lm}\,
  \vec{S}_{l} \cdot \vec{S}_{m}.
\end{equation}

Based on the semi-classical treatment\cite{Halilov98},
the energy
of spin-waves for a wave vector $\vec{q}$ can be obtained
as eigenvalues of the matrix in which the
$(i,j)$ component is
\begin{align}
  F_{ij}(\vec{q})
  &=
  2\delta_{i,j}\sum_k S_k \tilde{J}_{i,k}(\vec{0})
  -
  2S_i \tilde{J}_{i,j}(\vec{q}) \\
  &=
  2\delta_{i,j}\sum_k J_{i,k}(\vec{0})/S_i
  -
  2 J_{i,j}(\vec{q}) / S_j.
\end{align}
We obtain the spin-wave dispersion by diagonalizing this matrix
using the values of $J$ calculated by Eq.~\eqref{Eq:Liechtenstein_Formula_in_q}.

We calculate the spin-wave stiffness by fitting a quadratic function
from the curvature of the lowest branch around the $\Gamma$ point.
To compare the RSA with the spatial method, the spin-wave stiffness
is also estimated by the following matrix, which is valid when 
$q \ll 1$:
\begin{align}
  F_{ij} (\vec{q})\simeq
  & \frac{2}{S_j N}
  \sum_a
    \left(
      \delta_{ij}
      \sum_k J_{(i,0)(k,a)}
      -
      J_{(i,0)(j,a)}
    \right)
      \notag \\ 
  & -
  \sum_a
    J_{(i,0)(j,a)}
    \left(
      i \vec{q} \cdot (\vec{R}_a-\vec{R}_0)
      +
      \frac{1}{2} |\vec{q} \cdot (\vec{R}_a-\vec{R}_0)|^2
    \right).
    \label{Eq:Spatial_D}
\end{align} 
However, this estimate is numerically unstable
because the Fourier interpolation confines the curve onto 
the given points and
errors in $F_{ij}(\vec{q})$ generate fictitious high harmonics,
which directly lead to errors in the curvature.
In contrast, the quadratic regression that we use
is more robust against the errors. 
In addition, we do not have to calculate the spatial $J_{(i,a)(j,b)}$
and do not need to consume memory to store them.
Although calculation of the convolution in Eq.~\eqref{Eq:Liechtenstein_Formula_in_q}
is straightforward, it is also possible 
to use the fast Fourier transform to calculate $F_{ij} (\vec{q})$
when there is enough memory.

These methods are applied to Nd$_2$Fe$_{14}$B [space group: P4$_2$/mnm (\#136)]
and $R$Fe$_{11}$Ti ($R$=Y, Nd, Sm). 
The latter systems are obtained by replacing one of the Fe(8i) sites with Ti 
in $R$Fe$_{12}$ having the ThMn$_{12}$ structure
[space group: I4/mmm (\#139)].
We adopt lattice parameters from previous studies
obtained by numerical optimization using
first-principles calculations.
The parameters for Nd$_2$Fe$_{14}$B are taken from
Ref.~\onlinecite{Tatetsu18},
and those for $R$Fe$_{11}$Ti are
from Ref.~\onlinecite{Harashima14b}.

%
%
\section{Results and Discussion}
\label{S:Results}
\begin{figure}
  \includegraphics[width=8cm]{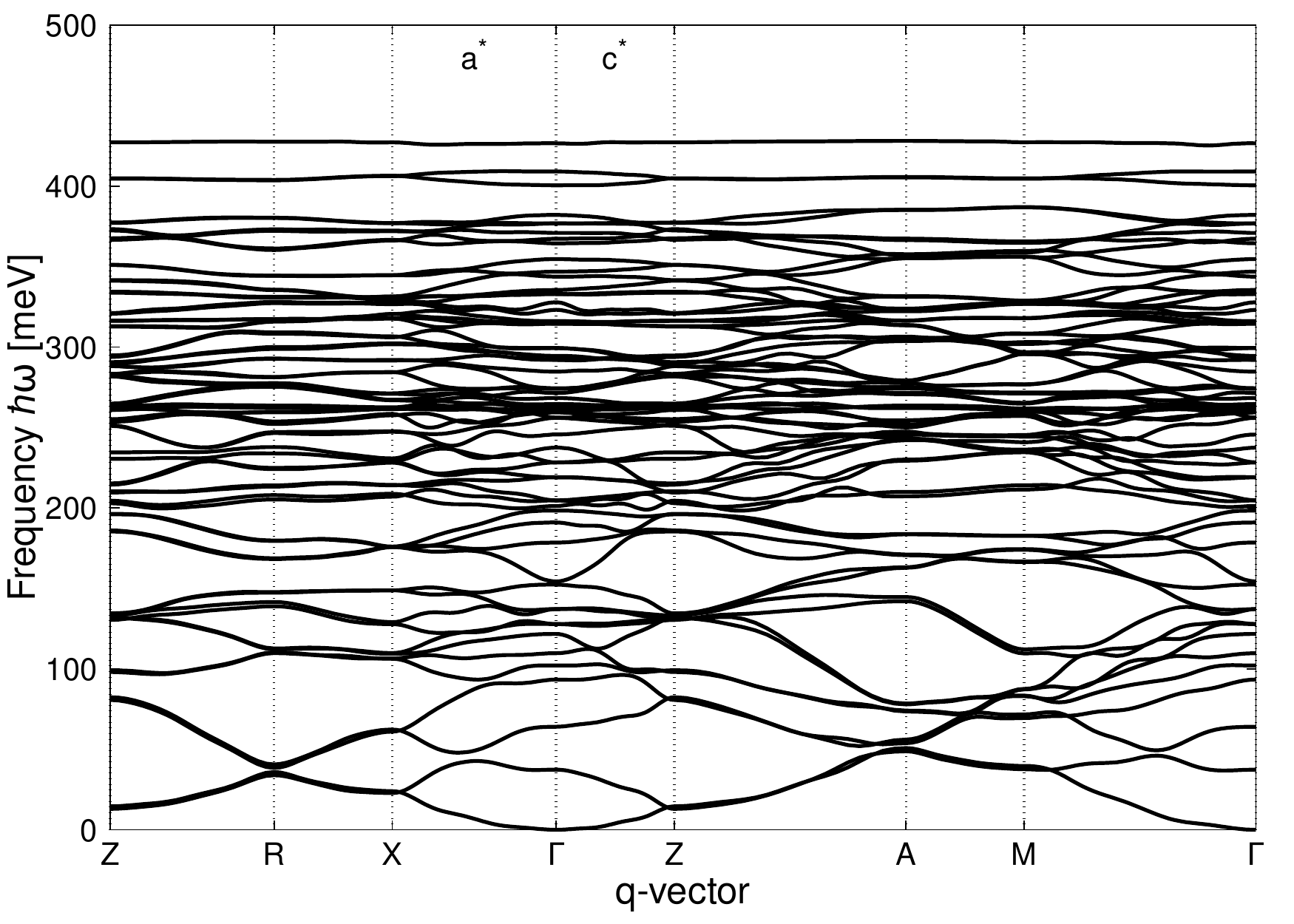}
  \caption{Spin-wave dispersion for Nd$_2$Fe$_{14}$B.}
  \label{spwave_Nd2Fe14B}
\end{figure}

The above-mentioned scheme enables us to calculate the spin-wave dispersion of 
Nd$_2$Fe$_{14}$B, which contains 68 atoms in a unit cell.
Figure~\ref{spwave_Nd2Fe14B} 
shows the calculated results. 
Although this scheme can give
branches associated with high excitation energy,
we should focus on low-energy excitations 
when comparing these results with experimental ones 
because 
the Heisenberg model we used in
the calculation is not valid for describing
excitations with high energy transfer. 
We hereafter discuss the lowest branch 
of spin waves
around the $\Gamma$ point 
through its spin-wave stiffness.

Table \ref{D_table} shows the spin-wave stiffness $D$
for Nd$_2$Fe$_{14}$B.
The curvature of the lowest branch around 
the $\Gamma$ point along the $a^*$ axis ($D_a$),
along the $b^*$ axis ($D_b$)
and along the $c^*$ axis ($D_c$) are also shown
in the table.
The spin-wave stiffness is converted to exchange stiffness by the following equation:\cite{Fukazawa19}
\begin{equation}
  A_{i'} = \rho D_i / 4 \quad [(i,i')=(a,x), (b,y), (c,z)],
\end{equation}
where $\rho$ is the number of Bohr magnetons per unit-cell volume.

The exchange stiffness $A$
and the components in the $x$-, $y$- and
$z$-direction ($A_x$, $A_y$ and $A_z$, respectively) are
shown in Table \ref{A_table}.
These results are within the range of experimental values by 
Ono et al. (6.6 pJ/m) and Naser et al. (18 pJ/m)
for Nd$_2$Fe$_{14}$B.

In our results,
the anisotropy ratio $A_z/A_x$ of the exchange stiffness
is approximately 1.1. Our prediction
is much more isotropic than Monte Carlo simulations
with ab initio modeling,
which gave $A_z/A_x \sim 0.8$
at the lowest temperature.\cite{Toga18,Gong20}
This deviation can be attributed to their use of 
spatial $J_{ij}$. Figure \ref{D_jtabl} shows the anisotropy ratio
calculated by diagonalizing the approximate $F_{ij}$ 
of Eq.~\eqref{Eq:Spatial_D} while varying the cutoff on
the $i$--$j$ distance.  At the spatial cutoff of 3.5 {\AA, which is the value
used in Ref.~\onlinecite{Toga18}, the value of $A_x$ becomes larger than $A_z$
and the anisotropy is inversely overestimated due to the artificial cutoff. 
Note that RSA does not have this kind of spatial cutoff.
\begin{figure}
  \includegraphics[width=8cm]{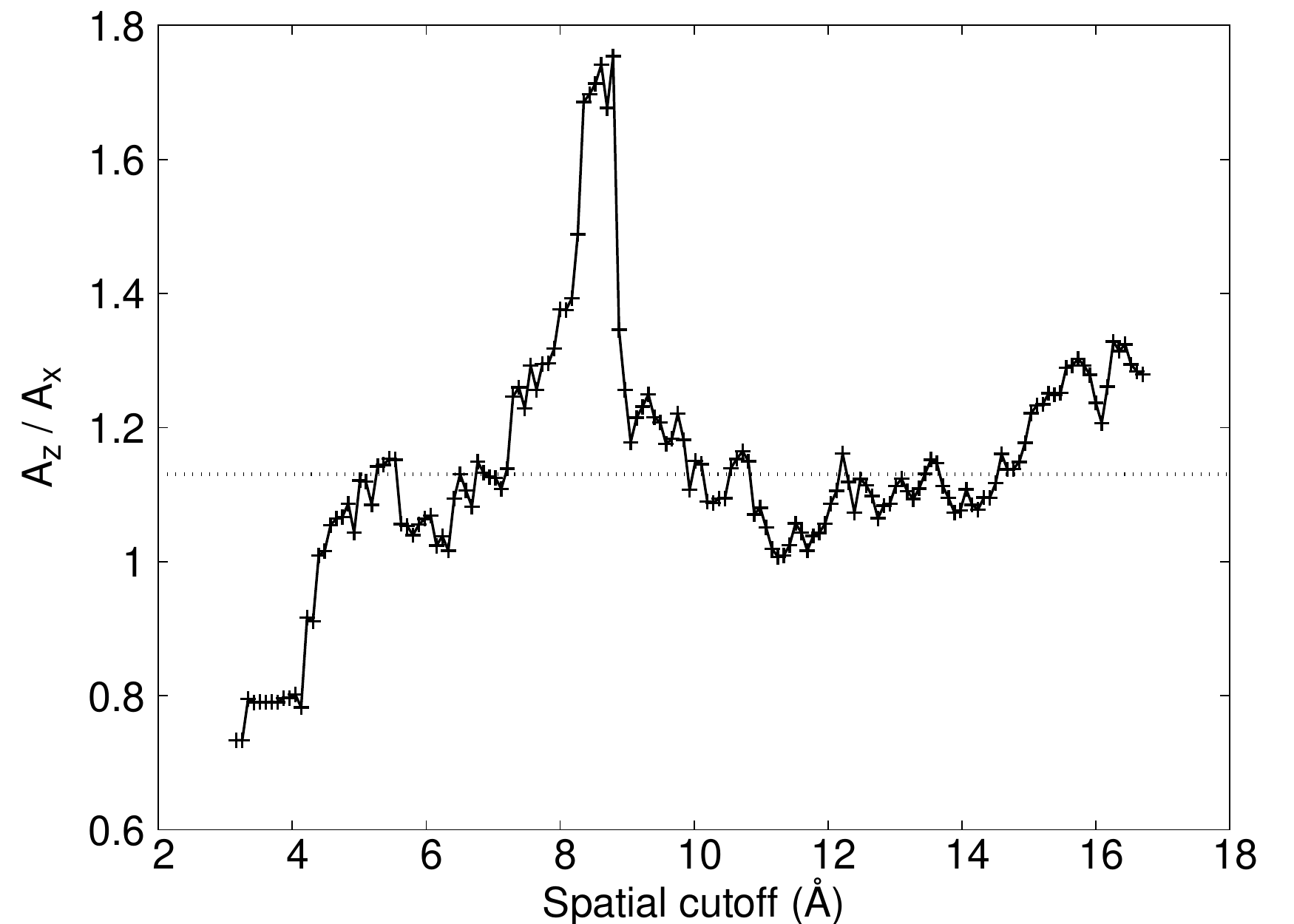}
  \caption{Anisotropy in the exchange stiffness for Nd$_2$Fe$_{14}$B
  calculated by Eq.~\eqref{Eq:Spatial_D}.
  The horizontal axis shows the spatial cutoff 
  for $J_{ij}$. 
  The horizontal dotted line shows the anisotropy ratio
  obtained by the RSA calculation.}
  \label{D_jtabl}
\end{figure}
Our results using RSA support the experimental observation by 
Mayer et al.\cite{Mayer92} that there are no
identifiable differences between the dispersion
along the $a^*$ and $c^*$ axis
although they did not quantitatively estimate the isotropy.
\begin{table}
 \caption{Values of spin-wave stiffness $D$
 for Nd$_2$Fe$_{14}$B, $R$Fe$_{11}$Ti, and SmFe$_{12}$. 
 $D_a$ is the curvature along the $a^*$ axis of the bottom branch
 around the $\Gamma$ point,
 $D_b$ along the $b^*$ axis, and $D_c$ along the $c^*$ axis.
 The values are in units of meV\AA$^2$.
 \label{D_table}}
 \begin{tabular}{rcccl}
  \hline
  \hline
  Formula  & $D_a$ & $D_b$ & $D_c$ & $D=\sqrt[3]{D_aD_bD_c}$\\
  \hline
  Nd$_2$Fe$_{14}$B & 190 & 190 & 215 & \quad 198 \\
  \hline
      YFe$_{11}$Ti & 117 & 167 & 249 & \quad 170 \\
     NdFe$_{11}$Ti & 178 & 158 & 262 & \quad 195 \\
     SmFe$_{11}$Ti & 172 & 168 & 271 & \quad 199 \\
     \hline
     SmFe$_{12}$\cite{Fukazawa19}   & 91.6& 91.6& 194 & \quad 118 (meV\AA$^2$)\\
  \hline
  \hline
 \end{tabular}
\end{table}
\begin{table}
\caption{Values of exchange stiffness $A$
for Nd$_2$Fe$_{14}$B, $R$Fe$_{11}$Ti and SmFe$_{12}$. 
$A_x$ is the exchange stiffness associated with
spin waves along the $x$ axis,
$A_y$ along the $y$ axis, and $A_z$ along $z$ axis.
The values are in units of pJ/m.
\label{A_table}}
\begin{tabular}{rcccl}
 \hline
 \hline
 Formula  & $A_x$ & $A_y$ & $A_z$ & $A=\sqrt[3]{A_xA_yA_z}$\\
 \hline
 Nd$_2$Fe$_{14}$B & 10.3 & 10.3 & 11.7 & \quad 10.7 \\
 \hline
     YFe$_{11}$Ti & 5.7  & 8.1 & 12.1 & \quad 8.3 \\
    NdFe$_{11}$Ti & 10.0 & 8.9 & 14.7 & \quad 10.9 \\
    SmFe$_{11}$Ti & 8.6  & 8.4 & 13.5 & \quad 9.9 \\
    \hline
    SmFe$_{12}$\cite{Fukazawa19}   & 5.6 & 5.6 & 7.2 & \quad 11.9 (pJ/m)\\
 \hline
 \hline
\end{tabular}
\end{table}

We also show the values of stiffness for $R$Fe$_{11}$Ti 
in Tables \ref{D_table} and \ref{A_table}.
The Ti element is located at the Fe(8i) site 
of $R$Fe$_{12}$ that is next to the $R$ atom
along the $a$ axis, as in the notation of
Harashima et al.\cite{Harashima14b}
The unit cell is orthorhombic, and the curvatures
along the $a$ and $b$ axes do not coincide by symmetry.
In these $R$Fe$_{12}$ systems, the stiffness along 
the $c$ axis is significantly larger than along
the $a$ and $b$ axes, which has also been seen in 
Sm(Fe,Co)$_{12}$.\cite{Fukazawa19}

We previously reported in Ref.~\onlinecite{Fukazawa19} that 
distortion of the Bravais lattice 
has an effect on the anisotropy of the spin-wave 
stiffness even when the values of $J_{ij}$ are fixed.
In the case of the transformations $a \rightarrow \alpha a$
and $c \rightarrow \gamma c$,
the $D_c / D_a (\equiv R)$ value is transformed to $(\gamma / \alpha)^2 R$ 
by the distortion.
This must not be confused with taking a supercell,
which involves band folding but does not change the curvatures.
It is noteworthy that the conventional cell of Nd$_2$Fe$_{14}$B
stacked 3 times along the $c$ axis ($a=8.791$ {\AA}, $c=36.42$ {\AA}, $D_c/D_a=1.1$) has
a similar length to the conventional cell of SmFe$_{12}$ 
stacked 8 times along the $c$ axis ($a=8.533$ {\AA}, $c=37.45$ {\AA}, $D_c/D_a=1.3$).
The lengths of SmFe$_{12}$ become identical to Nd$_2$Fe$_{14}$B
when a tetragonal distortion 
of $\alpha = 1.03$ and $\gamma = 0.97$
is applied to SmFe$_{12}$.
This distortion 
multiplies the factor of 
$(\gamma / \alpha)^2 = 0.89$ to
the anisotropy ratio, $D_c / D_a$,
and makes the ratio more isotropic: $R=1.2$, 
which corresponds to Nd$_2$Fe$_{14}$B.

To visualize the anisotropy, we calculate
angular integrated spin-wave dispersions
according to Eq.~\eqref{eq:I_q_E_2} in Appendix \ref{Appendix_I_E_q}.
Figure \ref{spwave_averaged} shows the density of 
states for Nd$_2$Fe$_{14}$B and $R$Fe$_{11}$Ti
as functions of the absolute value, $q$, of the wave vector
and the energy, $E$.
These dispersions correspond to diffraction
of polycrystalline and powder samples.
\begin{figure}
  \includegraphics[width=8cm]{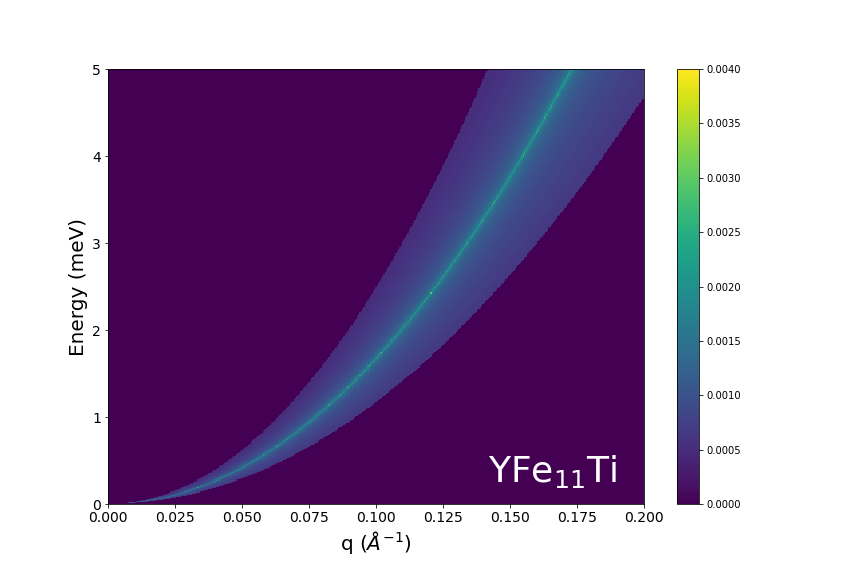}
  \includegraphics[width=8cm]{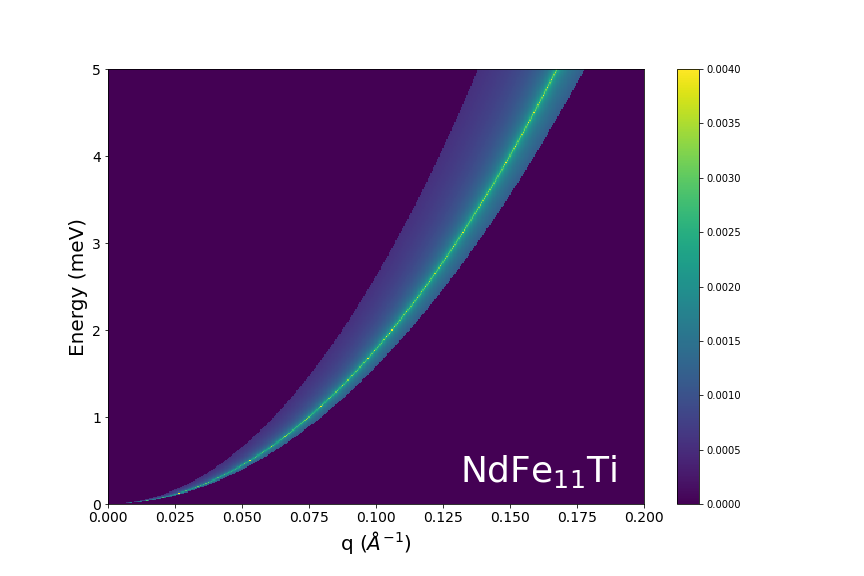}
  \includegraphics[width=8cm]{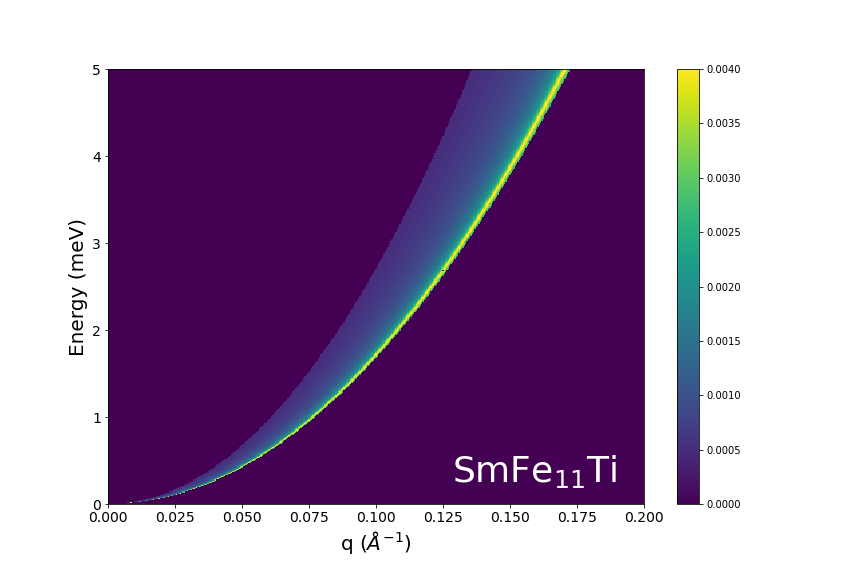}
  \includegraphics[width=8cm]{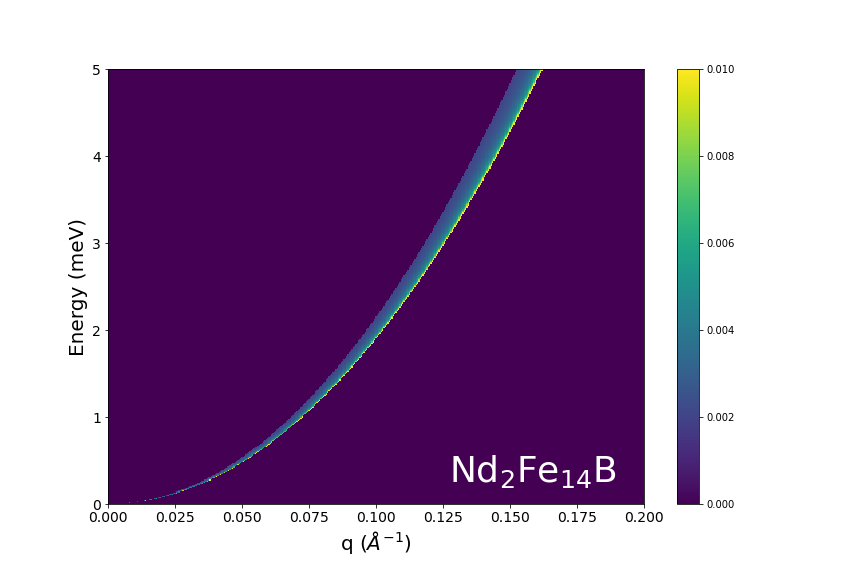}
  \caption{\label{spwave_averaged}
  (Color online)
  Angular-integrated spin-wave dispersion
  for $R$Fe$_{11}$Ti and Nd$_2$Fe$_{14}$B.
  These color maps show the density of states $I(q,E)$
  of Eq.~\eqref{eq:I_q_E_2} where the horizontal axis 
  denotes the absolute value of a wave vector, $q$, and the 
  vertical axis is the energy, $E$. }
\end{figure}
As mentioned in Appendix \ref{Appendix_I_E_q}, the lower edge of
the finite-DOS region is given 
by $E=D_{\min} q^2$ and the upper edge by $E=D_{\max} q^2$,
where
$D_{\min}=\min(D_a, D_b, D_c)$ and
$D_{\max}=\max(D_a, D_b, D_c)$.
Therefore, the anisotropy of the spin-wave stiffness
is reflected in the smearing of the dispersion in
these figures.
The smearing in Nd$_2$Fe$_{14}$B is much smaller than
the measurement errors in the neutron scattering
experiment by Ono et al.\cite{Ono14}.
Hawai et al. recently performed inelastic neutron diffraction
experiments for $R$Fe$_{11}$Ti.\cite{Hawai20}
Although anisotropy could not be seen in their data
due to uncertainties in the experiment,
further development of neutron scattering techniques will hopefully 
resolve the existence of the anisotropy.

\section{Conclusion}
\label{S:Conclusion}
We presented 
the spin-wave dispersion for Nd$_2$Fe$_{14}$B
and $R$Fe$_{11}$Ti obtained using
first-principles calculations.
Our recently developed RSA, an accelerated scheme for
spin-wave dispersion that is based on
the Liechtenstein formula and its use in reciprocal space,
enabled these calculations in practical time. 
We discuss the anisotropy of the exchange stiffness
in Nd$_2$Fe$_{14}$B. 
Our results suggest that 
the anisotropy in the exchange stiffness of Nd$_2$Fe$_{14}$B
is much smaller than the values obtained by
previous Monte Carlo simulations.
We also demonstrated that the spatial cutoff can be a crucial factor in the
precision when estimating the anisotropy.
However, $R$Fe$_{11}$Ti has a strong 
anisotropy in our estimation of the spin-wave stiffness.
We also showed the angular integrated dispersion of
the spin-waves, which corresponds to diffraction
of polycrystalline and powder samples. 
We pointed out that the anisotropy can be seen 
as smearing of the $E$--$q$ dispersion,
and this is comparable with neutron scattering 
experiments.

\section*{Acknowledgment}
We are grateful to Takafumi Hawai, Kanta Ono, and Nobuya Sato
for their advice and fruitful discussions.
We acknowledge support
from the Elements Strategy Initiative Center for Magnetic Materials (ESICMM), 
Grant Number JPMXP0112101004, under the auspices of the Ministry of Education, Culture, Sports, Science and Technology (MEXT).
This work was also supported by MEXT through the ``Program for Promoting Researches on the Supercomputer Fugaku''
(DPMSD).
The computation was partly conducted using the facilities of the Supercomputer Center at
the Institute for Solid State Physics, University of Tokyo,
and the supercomputer of the Academic Center for Computing and Media Studies (ACCMS), Kyoto University. 
This research also used computational resources of the K computer provided
by the RIKEN Advanced Institute for Computational Science
through the HPCI System Research project (Project ID:hp170100). 

\appendix
\section{Convergence with respect to the number of $k$ points in Fe and Co}
We examine the convergence of the calculation for bcc-Fe and hcp-Co with respect to the number of $k$-points.
Figure~\ref{spwave_FeCo} shows the spin-wave dispersion obtained 
by our calculation. The solid lines are for the calculations with 
the largest number of $k$-points.
The overall behavior of the curve agrees
with previous studies\cite{Halilov98,Pajda01}.
The dashed curves show the results with a smaller number
of $k$-points, and the dash-dotted curves
with an even smaller number.

In the case of Fe, we need a large number of $k$-points
to reproduce the description of the spin-wave dispersion
with $18 \times 18 \times 18$ $k$-points (solid).
The dot-dashed curve shows the dispersion
with $8 \times 8 \times 8$ $k$-points (dot-dashed),
which is the cheapest calculation among the three
and has a significant deviation from the solid curve.
With $12 \times 12 \times 12$ $k$-points (dashed),
some features such as the depth of the dip on
the $\Gamma$--H line are missing. 
However, we confirmed that 
the curves with $14 \times 14 \times 14$ $k$-points 
appear much closer to the solid curve.
In the case of Co, convergence is much faster
owing to the smallness of the Brillouin zone, and
$9 \times 9 \times 9$ $k$-points seem to be enough.
\begin{figure}
  \includegraphics[width=8cm]{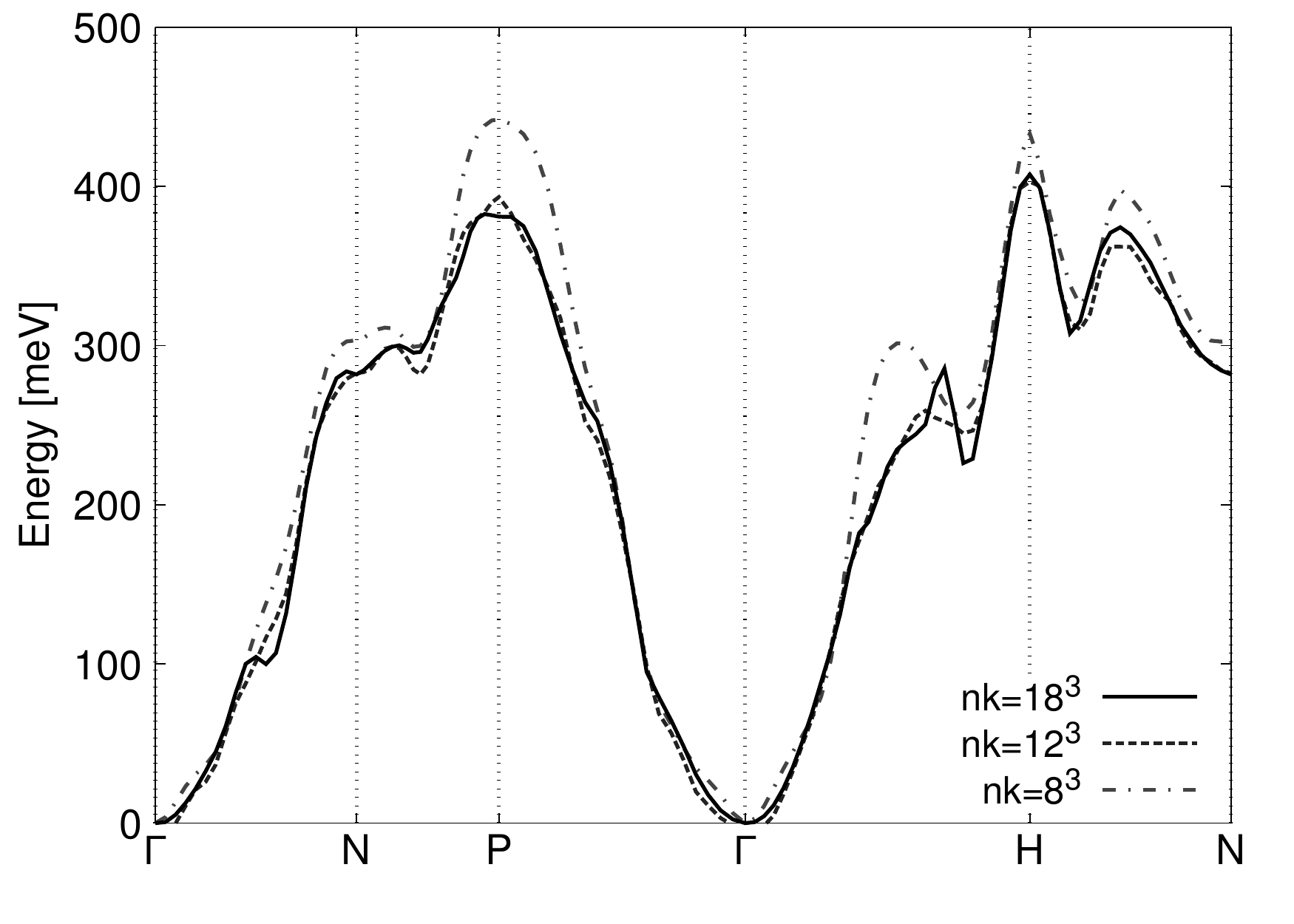}
  \includegraphics[width=8cm]{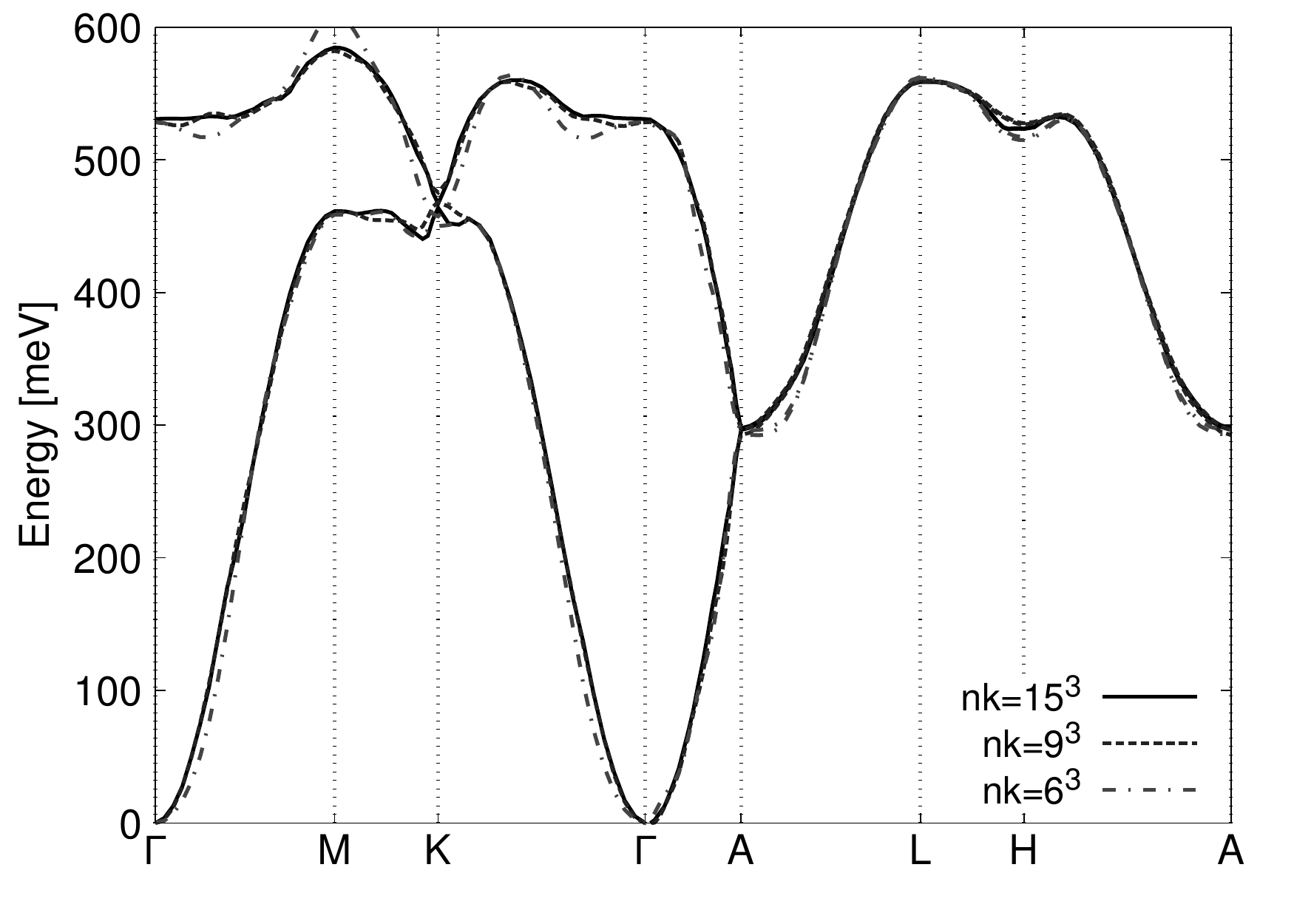}
  \caption{Spin-wave dispersion in (Top) bcc-Fe and (Bottom) hcp Co.
  The solid lines are from the calculations with the largest number
  of $k$-points (nk): $18\times 18 \times 18$ for Fe;
  $15 \times 15 \times 15$ for Co.
  The dashed lines and the dash-dot lines are for
  smaller nk.}
  \label{spwave_FeCo}
\end{figure}

Figure~\ref{convergence_Fe} shows the convergence of 
the spin-wave stiffness---the curvature around the 
$\Gamma$ point in the lowest branch---with respect to
the number of $k$-points.
The filled circles show the results when the SCF potential 
was obtained with the same number of $k$-points
as used in the spin-wave calculation.
The spin-wave stiffness appears well-behaved as a function
of the number of $k$-points and converges rapidly.
\begin{figure}
  \includegraphics[width=8cm]{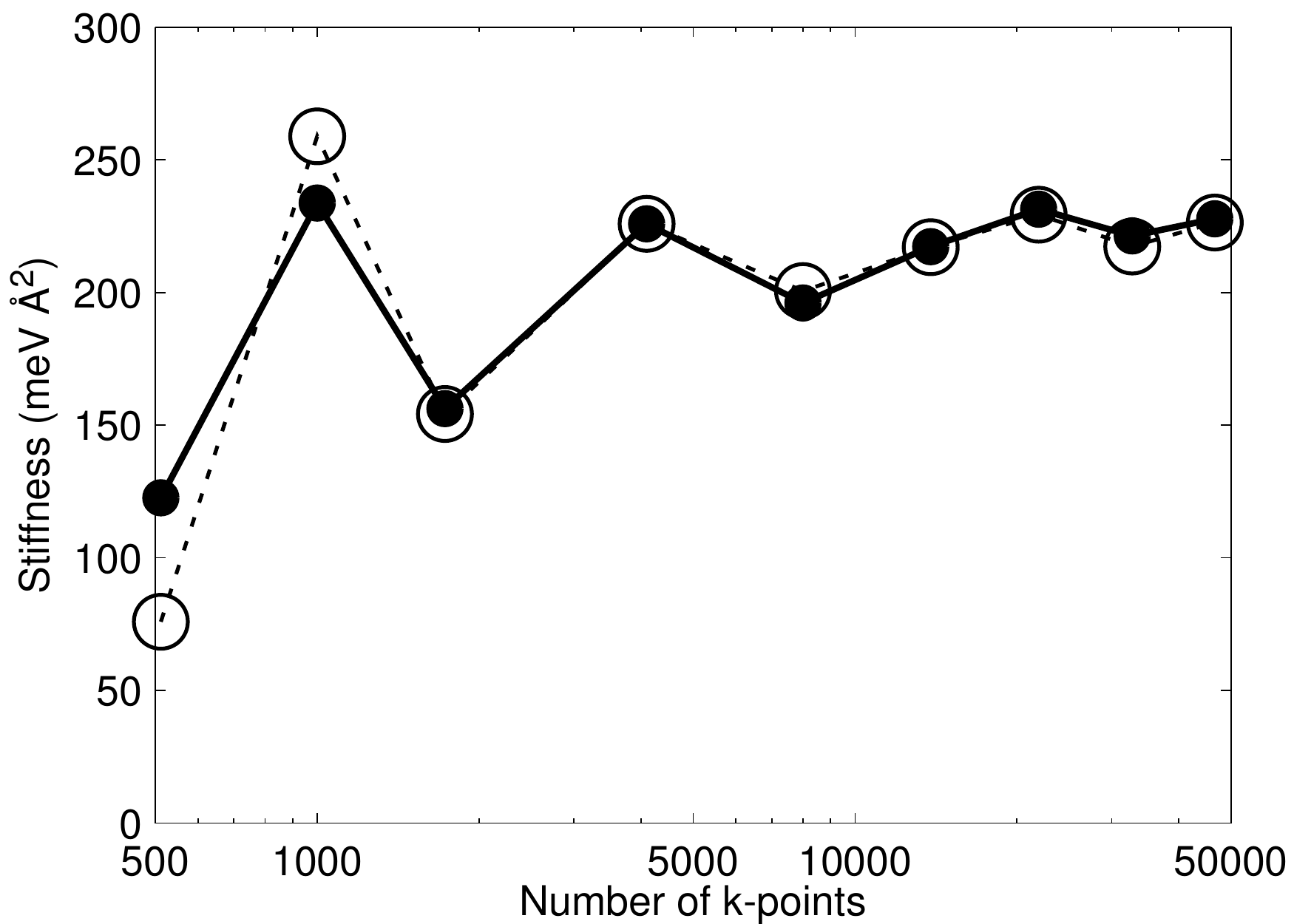}
  \caption{Convergence of the spin-wave stiffness
  with respect to
  the number of the $k$-points (nk) where 
  nk for SCF is
  the same as the spin-wave calculation (filled circle),
  and where nk is fixed to 1000 (open circle).
  Note that the horizontal axis is in log scale.
  }
  \label{convergence_Fe}
\end{figure}
We also show the results for the SCF potential fixed to 
that obtained with $10 \times 10 \times 10$ $k$-points
as open circles.
By comparison,
we observe that
the quality of the potential does not have much effect
on the spin-wave stiffness.

\section{$E$--$q$ dispersion of spin-waves}
\label{Appendix_I_E_q}
To compare the theoretical results with
experimental results for polycrystalline and powder samples,
we perform angular integration of
the following energy dispersion:
\begin{equation}
  E = D_a q_a^2 + D_b q_b^2 + D_c q_c^2
  \quad (0 \leq D_a \leq D_b \leq D_c),
  \label{Eq:starting_dispersion}
\end{equation}
which is adequate when the low-energy excitations are
of interest. The inequality for the coefficients does not
deteriorate the generality because it is always satisfied
by exchanging the labels of axes. 

It is obvious that there are states only when
$D_a q^2 \leq E \leq D_c q^2$ holds.
We divide the problem into 3 cases:
(i)   $D_a q^2 \leq E < D_b q^2$,
(ii)  $E = D_b q^2$, and 
(iii) $D_b q^2 < E \leq D_c q^2$.
In the case of (i),
by eliminating $q_a^2$ in $E$
with $q_a^2=q^2-q_b^2-q_c^2$,
we can obtain
\begin{gather}
  \frac{x^2}{\alpha^2}
  +
  \frac{y^2}{\beta^2}
  =
  \epsilon^2
  \quad (0 < \alpha \leq \beta),
  \label{Eq:std_form_1}\\
  z^2 = q^2 - x^2 -y^2, \label{Eq:std_form_2}
\end{gather}
with
\begin{gather}
  \alpha = \sqrt{\frac{1}{D_c-D_a}}, \quad 
  \beta  = \sqrt{\frac{1}{D_b-D_a}}, \quad 
  \epsilon = \sqrt{E-D_a q^2}, \\
  x=q_c, \quad y=q_b, \quad z=q_a. \label{Eq:q_x_transform_i}
\end{gather}
In the case of (ii), we can immediately obtain
\begin{equation}
  (D_b-D_a)q_a^2 = (D_c-D_b)q_c^2
\end{equation}
by eliminating $q_b$ in $E$.
In the case of (iii),
by eliminating $q_c^2$ in $E$
with $q_c^2=q^2-q_a^2-q_b^2$,
we can obtain 
the same set of equations as
Eq.~\eqref{Eq:std_form_1} and Eq.~\eqref{Eq:std_form_2}
with 
\begin{gather}
  \alpha = \sqrt{\frac{1}{D_c-D_a}}, \quad 
  \beta  = \sqrt{\frac{1}{D_c-D_b}}, \quad 
  \epsilon = \sqrt{D_c q^2 - E}, \\
  x=q_a, \quad y=q_b, \quad z=q_c. \label{Eq:q_x_transform_iii}
\end{gather}
Therefore,
we can express
the equations for 
the possible sets of $q_a, q_b, q_c$ by
a single set of equations---Eq.~\eqref{Eq:std_form_1}
and Eq.~\eqref{Eq:std_form_2}---except for
the singular point at which (ii) $E=D_b q^2$ holds.

When considering the density of states of spin-waves,
we can disregard case (ii) because it has a measure of zero
in the integral with respect to energy.
Let us consider a solution
$\vec{P}=(x,y,z)$ 
of \eqref{Eq:std_form_1}
and \eqref{Eq:std_form_2}
in the $z>0$ region.
By the following transform:
\begin{align}
  x &= \epsilon \alpha \cos \theta, \\
  y &= \epsilon \beta \sin \theta,
\end{align}
this vector can be expressed as follows:
\begin{equation}
  \vec{P}
  =
  \left(
    \epsilon \alpha \cos \theta,
    \epsilon \beta \sin \theta,
    \sqrt{q^2-\epsilon^2(\alpha^2 \cos^2 \theta + \beta^2 \sin^2 \theta)}
  \right).
\end{equation}
When $\Delta E$ and $\Delta q$ are infinitesimal numbers,
the number of states within the energy range
from $E$ to $E + \Delta E$
and within the momentum range from $q$ to $q + \Delta q$ 
can be expressed as
\begin{equation}
  I(q, E) \Delta E \Delta q
  =
  \frac{2V}{8\pi^3}
  \int^{\pi}_{-\pi}
  d\theta
  \left|
  \frac{\partial (x,y,z)}{\partial (q, E, \theta)}
  \right|
  \Delta E \Delta q
  \label{eq:I_q_E}
\end{equation}
where $V$ denotes the system volume.
The factor of two is necessary because there is
another solution $(x,y,-z)$ in $z < 0$.
By straightforward calculation, we can obtain an expression
with the complete elliptic integral of the first kind, $K(k)$,
as follows:
\begin{equation}
  I(q,E)
  =
  \frac{V}{\pi^3}\,
  \alpha \beta
  \frac{q}{\sqrt{q^2-\epsilon^2\alpha^2}}\,
  K\left(
      \sqrt{
        \frac{(\beta^2-\alpha^2)\epsilon^2}{q^2-\epsilon^2\alpha^2}
      }
    \right),
    \label{eq:I_q_E_2}
\end{equation}
An outline of this is shown in Fig.~\ref{Outline_I_q_E}.
Although we can also write 
the elliptic modulus,
$k=\sqrt{ {(\beta^2-\alpha^2)\epsilon^2}/{(q^2-\epsilon^2\alpha^2)}}$,
in terms of the variables in Eq.~\eqref{Eq:starting_dispersion},
the expressions for cases (i) and (iii) are different:
\begin{align}
  k^2&={(\beta^2-\alpha^2)\epsilon^2}/{(q^2-\epsilon^2\alpha^2)}
  \notag \\
  &=
  \begin{dcases}
    \frac{D_c-D_b}{D_c q^2 - E} \frac{E-D_a q^2}{D_b - D_a}
    & \text{Case (i)} \\
    \frac{D_b-D_a}{E - D_a q^2} \frac{D_c q^2 - E}{D_c - D_b}
    & \text{Case (iii)}
  \end{dcases},
\end{align}
with which one can observe $0 \leq k^2 < 1$.
When $E$ is fixed, $k$ is a monotonically increasing function of $q$ 
in case (i) while $k$ is monotonically decreasing in case (iii).
Because $K(k)$ is also a monotonically increasing function,
we observe
\begin{equation}
  I(E) \leq \lim_{E\rightarrow D_b q^2} I(E)
  = I(D_b q^2).
\end{equation}
Therefore, the density of states is
most intense along the line of $E=D_b q^2$.
\begin{figure}
  \includegraphics[width=4.3cm]{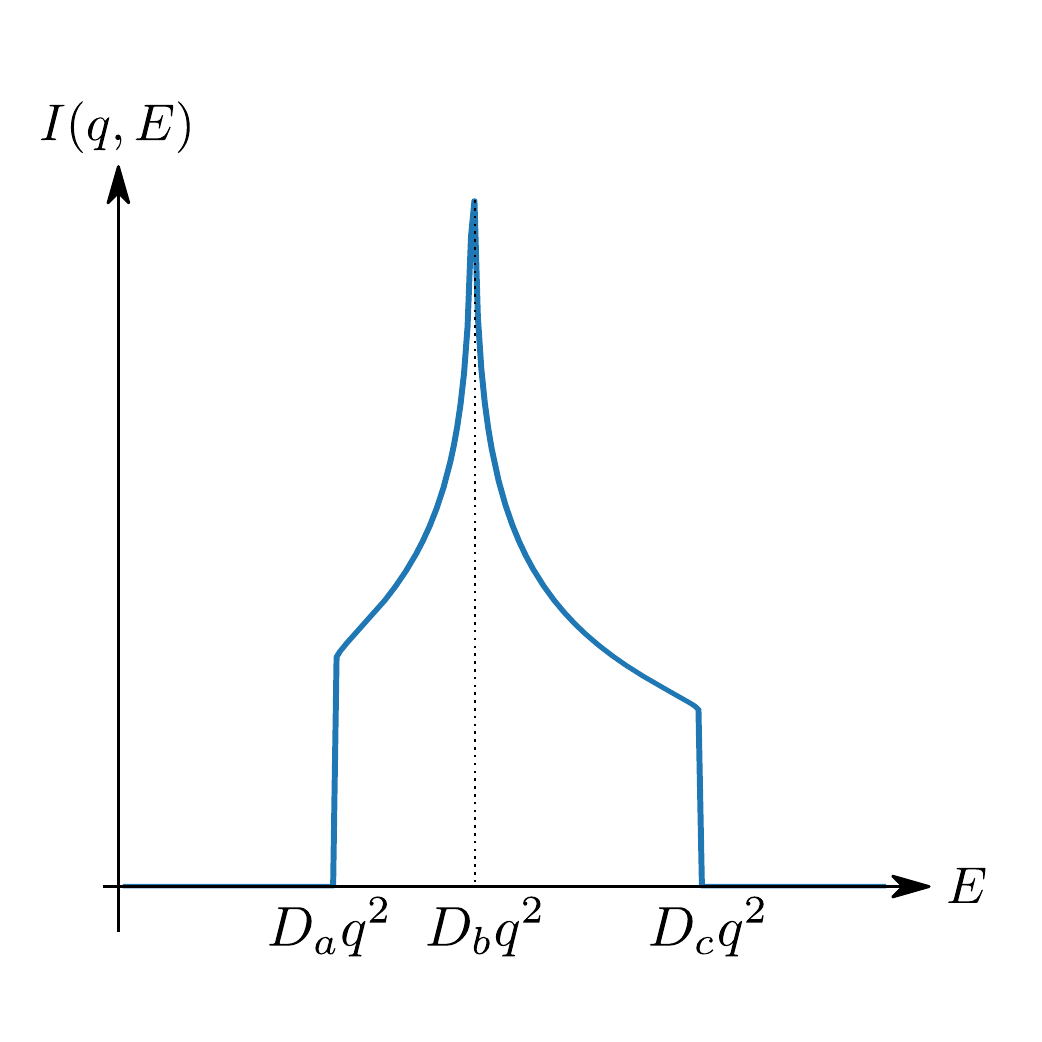}
  \hspace{-2mm}
  \includegraphics[width=4.3cm]{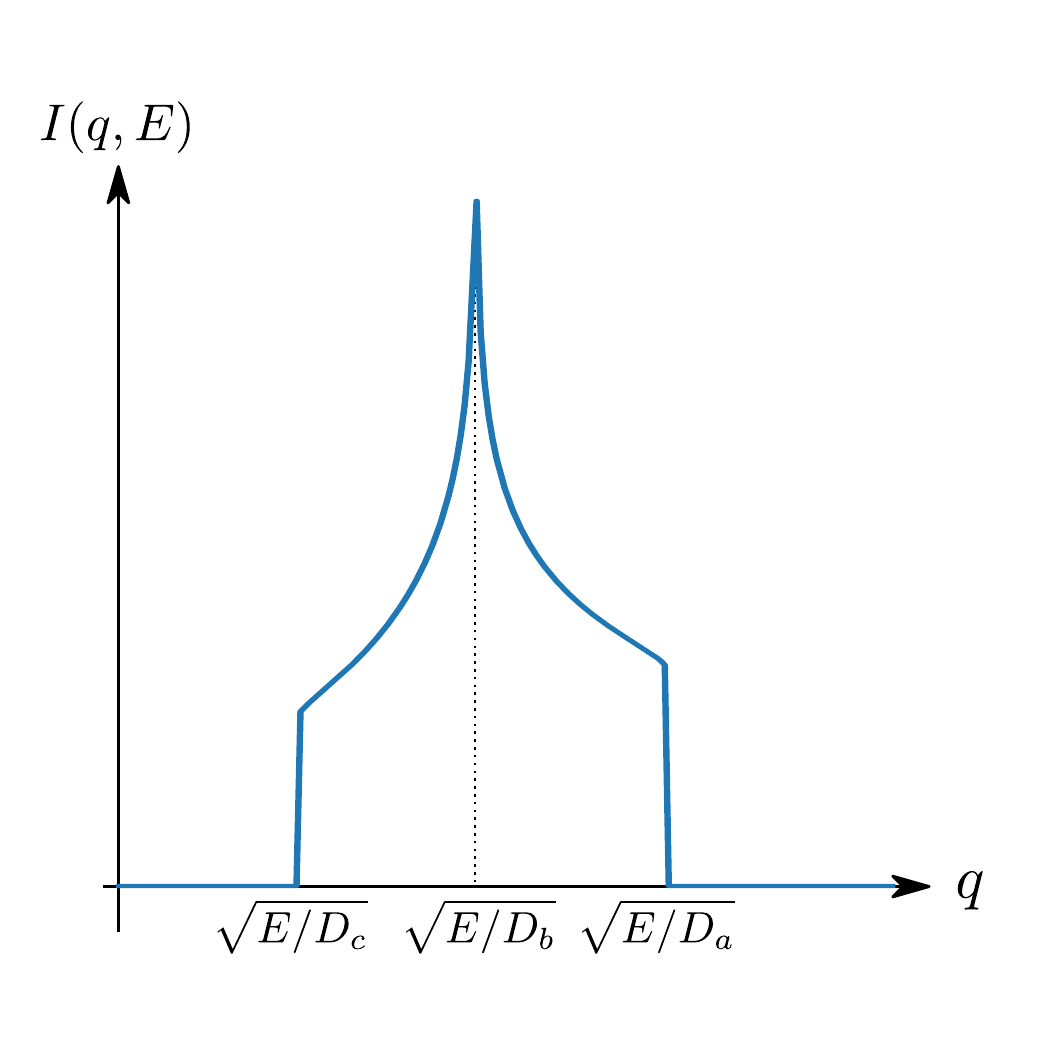}
  \caption{Outlines of $I(q,E)$ in Eq.~\eqref{eq:I_q_E_2}
  where (Left) $q$ is fixed and (Right) $E$ is fixed.}
  \label{Outline_I_q_E}
\end{figure}

\bibliography{spinwave2}
\end{document}